\documentstyle[11pt,aaspp4]{article} 
\begin{document}
\title{Dissipative Strong-Field Electrodynamics} 
\author{Andrei Gruzinov}
\affil{CCPP, Physics Department, New York University, 4 Washington Place, New York, NY 10003}

\begin{abstract}

A dissipative Lorentz-covariant Ohm's law which uses only the electromagnetic degrees of freedom is proposed. For large conductivity, Maxwell equations equipped with this Ohm's law reduce to the equations of Force-Free Electrodynamics (FFE) with small dissipative corrections,  but only in the regions where the ideal FFE 4-current is space-like. This might indicate that the pulsar emission comes primarily from the magnetic separartrix.

\end{abstract}

\section{Introduction}

Force-Free Electrodynamics (FFE) supposedly describes large-scale electromagnetic fields in the magnetospheres of powerful Crab-like pulsars (for weaker pulsars, FFE probably applies only in the high-field region close to the neutron star). Since FFE uses an ideal dissipationless Ohm's law, it cannot explain the pulsar emission. Here we add dissipation to FFE. The resulting system is no longer force-free, we call it Strong-Field Electrodynamics (SFE).

In \S2 we recall the definition of FFE and define SFE. SFE is derived in \S3. We discuss how SFE works (and when it fails!) in \S4. 

While pulsars (and maybe accreting black holes) are the true motivation for FFE and SFE, we do not discuss the pulsar astrophysics in this paper. The classical papers Goldreich \& Julian (1969) and Rutherman \& Sutherland (1975) give a good introduction.

\section{Force-Free and Strong-Field Electrodynamics }

FFE describes electromagnetic fields of special geometry (for derivation and applicability of FFE see Gruzinov 2007). The fields create electron-positron plasma by some pair production mechanism, but the plasma is described only implicitly, by postulating the following ideal (that is dissipationless) Ohm's law 
\begin{equation}\label{ohm}
{\bf j}={({\bf B}\cdot \nabla \times {\bf B}-{\bf E}\cdot \nabla \times {\bf E}){\bf B}+(\nabla \cdot {\bf E}){\bf E}\times {\bf B} \over B^2}.
\end{equation}
The electromagnetic field is evolved according to Maxwell equations, 
\begin{equation}\label{maxwell}
\partial _t{\bf B} =-\nabla \times {\bf E},~~~ \partial _t{\bf E}=\nabla \times {\bf B}-{\bf j}.
\end{equation}
It is assumed that the initial electric field is everywhere smaller than and perpendicular to the magnetic field
\begin{equation}\label{inv}
{\bf E}\cdot {\bf B}=0.
\end{equation}
The FFE Ohm's law was designed (Gruzinov 1999) to maintain the constraint (\ref{inv}) and to provide the so-called force-freedom 
\begin{equation}\label{force}
\rho {\bf E}+{\bf j}\times {\bf B}=0,
\end{equation}
where $\rho \equiv\nabla \cdot {\bf E}$ is the charge density. 

FFE is Lorentz-covariant. The 3+1 formulation (equations (\ref{ohm}), (\ref{maxwell})) makes it obvious that one can formulate the Cauchy problem in FFE, but hides the Lorentz-covariance. An explicitly covariant form of FFE is given by Maxwell, force-free, and vanishing invariant equations
\begin{equation}\label{covFFE}
\partial _\nu F^{\mu \nu}=-j^\mu , ~~~ F^{\mu \nu}j_\nu =0, ~~~ F^{\mu \nu}\tilde{F}_{\mu \nu}=0,
\end{equation}
where $F$ is the electromagnetic field tensor, $j$ is the 4-current, $\tilde{F}$ is the dual tensor.

It appears that magnetospeheres of Crab-like pulsars can be described by FFE (Gruzinov 2007). This allows to calculate the pulsar magnetosphere, understand the singularity structure, and find the spin-down power (Contopoulos, Kazanas, Fendt 1999, Gruzinov 2005, 2006, Spitkovsky 2006). 

Being ideal, FFE makes no predictions regarding the pulsar emission. Except that FFE predicts the existence and stability of a singular current layer along the magnetic separatrix. For any reasonable dissipation, one expects that the separatrix dominates the pulsar emission.

To make progress on the pulsar emission problem and to resolve the singularity one needs to add dissipation to FFE. This can be done by dropping the ${\bf E}\cdot {\bf B}=0$ assumption and replacing the ideal FFE Ohm's law (\ref{ohm}) by the following dissipative Ohm's law
\begin{equation}\label{ohmm}
{\bf j}={\rho {\bf E}\times {\bf B} +(\rho ^2+\gamma ^2\sigma ^2E_0^2)^{1/2}(B_0{\bf B}+E_0{\bf E})\over \gamma ^2(B_0^2+E_0^2)}.
\end{equation}
The $\gamma$ symbol and the scalars in the above are: 

(i) $\gamma$ is the Lorentz factor of a certain frame where ${\bf E}$ becomes parallel to ${\bf B}$:
\begin{equation}\label{gamma}
\gamma \equiv {1\over \sqrt{1-\beta^2}}, ~~~ {\beta \over 1+\beta ^2}\equiv {|{\bf E}\times {\bf B}|\over B^2+E^2},
\end{equation}

(ii) the scalars $B_0$ and $E_0$ represent the field invariants:
\begin{equation}\label{B0E0}
B_0^2-E_0^2\equiv {\bf B}^2-{\bf E}^2, ~~~ B_0E_0\equiv {\bf E}\cdot {\bf B}, ~~~E_0\geq 0,
\end{equation}

(iii) the conductivity scalar $\sigma$ is an arbitrary function of the field invariants:
\begin{equation}\label{sigma}
\sigma=\sigma(B_0,E_0).
\end{equation}

The system (\ref{maxwell}), (\ref{ohmm}) gives an evolutionary formulation of SFE. The covariant formulation is
\begin{equation}\label{covSFE}
\partial _\nu F^{\mu \nu}=-j^\mu , ~~~ B_0F^{\mu \nu}j_\nu =E_0\tilde{F}^{\mu \nu}j_\nu , ~~~ F^{\mu \nu}j_\nu F_{\mu \lambda}j^\lambda =\sigma^2E_0^4.
\end{equation}

\section{SFE Ohm's law}

The SFE Ohm's law is dissipative, ${\bf E}\cdot {\bf j}\geq 0$ -- the electromagnetic energy decreases in any frame. 

In the high conductivity limit, SFE reduces to FFE in the following way. The current $\sigma E_0$ stays finite, while $\sigma \rightarrow \infty$ and $E_0\rightarrow 0$, so that ${\bf E}\cdot {\bf B}\rightarrow 0$. For ${\bf E}\cdot {\bf B}=0$, eq.(\ref{gamma}) gives $\beta =E/B$, and then eq.(\ref{ohmm}) becomes
\begin{equation}\label{ohmmm}
{\bf j}={wB_0{\bf B}+\rho {\bf E}\times {\bf B} \over B^2}.
\end{equation}
where $w\equiv \sqrt{\rho ^2+\gamma ^2\sigma ^2E_0^2}$. To sustain the ${\bf E}\cdot {\bf B}=0$ condition, the current $\sigma E_0$ picks such a value that eq.(\ref{ohmmm}) reduces the ideal Ohm's law (\ref{ohm}). 

Note however, that as the current $\sigma E_0$ changes from $0$ to $\infty$, the coefficient $w$ in the Ohm's law (\ref{ohmmm}) never goes to zero, $w\geq |\rho |$. This gives ${\bf j}^2-\rho ^2\geq 0$. Thus SFE reduces to FFE only in the regions where the FFE 4-current is space-like.

To explain the origin of the SFE Ohm's law, recall that the FFE Ohm's law can be derived as follows. At each event, one chooses some {\it good} frame, where ${\bf E}$ is parallel to ${\bf B}$, and postulates that in any {\it good} frame, the electric field vanishes and the current flows along the magnetic field. In SFE, one postulates the existence of the {\it best} frame obtained by an appropriate boost along the family of {\it good} frames, where ${\bf E}$ is parallel to ${\bf B}$, the current flows along ${\bf B}$, and the charge density vanishes. In the {\it best} frame, the current is equal to $\sigma E_0$. The Ohm's law (\ref{ohmm}) simply implements this procedure for assigning the current at each event.

The above derivation of the SFE Ohm's law explains why the local dissipative Ohm's law can exist only in the regions with space-like current. We might add the following microscopic plausibility argument. Any large electric field will move the positrons and electrons in opposite directions at the speed of light, thus creating a space-like current (or a null current if one of the species is missing).

\section{SFE solutions}
We used direct numerical simulations to see how the SFE works in some simple cases. When the corresponding ideal FFE problem has space-like currents, the SFE just provides well-behaved dissipative correction. But when one tries to describe the charge relaxation using SFE, the model breaks down.

{\it No charge:} Take a simple 1+1 dimensional problem with a well-behaved ideal FFE solution -- consider the electromagnetic field of the following form 
\begin{equation}\label{eg1}
{\bf B}=(0,B_y(t,x),B_z(t,x)), ~~{\bf E}=(0,E_y(t,x),E_z(t,x)).
\end{equation}
The FFE solutions are nonlinear waves, including standing shocks (some examples are given by Gruzinov 1999). The FFE current is space-like. As expected, the SFE damps the waves and spreads the shock. The solutions are well-behaved.

{\it Charge damping:} Now take 
\begin{equation}\label{eg2}
{\bf B}=(B_0,0,0), ~~{\bf E}=(E(t,x), 0,0).
\end{equation}
This field configuration has no FFE analog, so it might be interesting to see what the SFE will do to it. SFE gives a Lorentz-invariant equation 
\begin{equation}\label{1dc}
\dot{E} ^2-E'^2=\sigma ^2E^2.
\end{equation}
For smooth initial field, the evolution is well-behaved at first, but the maxima of $E$ develop finite-time singularities (infinite second derivative). This is not too surprising, since the corresponding FFE limit does not exist.

{\it Pulsars:} The problem with pulsars is that the magnetosphere contains the regions with both space-like and time-like currents (Gruzinov 2006). According to what we have now, we must describe the time-like current regions by FFE and the space-like current regions by SFE. While it remains unclear if such a description is at all possible, the indication is that the space-like current regions are the primary sites of dissipation and should therefore be the primary sites of pulsar emission. The strongest space-like current flows along the magnetic separatrix.

\acknowledgements
This work was supported by the David and Lucile Packard foundation.

\end{document}